\documentclass[aps,prl,reprint,showpacs,amsmath,amssymb]{revtex4-1}

\usepackage{braket}
\usepackage{graphicx}
\usepackage{natbib}
\usepackage{hyperref}
\usepackage[caption=false]{subfig}  
\usepackage[table]{xcolor}  

\newcommand{\nM}{n_{\rm M}}
\newcommand{\Wc}{W_{\rm c}}

\newcommand{\Wcn}{W_{\rm c}^{\rm 0}}

\newcommand{\ac}{a_{\rm c}}
\newcommand{\Ls}{L_{\rm s}}
\newcommand{\De}{D_{\rm e}}
\newcommand{\NP}{N_{\rm P}}
\newcommand{\ND}{N_{\rm D}}
\newcommand{\nF}{n_{\rm F}}

\newcommand{\Ft}{\tilde{F}}
\newcommand{\Et}{\tilde{E}}
\newcommand{\taus}{\tau_{\rm s}}

\newcommand{\taun}{\tau_{\rm 0}}
\newcommand{\rhotyp}{\rho_{\rm typ}}
\newcommand{\rhoav}{\rho_{\rm av}}
\newcommand{\etat}{\tilde{\eta}}
\newcommand{\Xs}{X_{\rm s}}

\newcommand{\XB}{X_{\rm B}}
\newcommand{\lB}{l_{\rm B}}
\newcommand{\varphis}{\varphi_{\rm s}}
\newcommand{\varphia}{\varphi_{\rm a}}

\newcommand{\epsF}{\varepsilon_{\rm F}}
\newcommand{\vF}{v_{\rm F}}
\newcommand{\EM}{E_{\rm M}} 
\newcommand{\NS}{N_{\rm S}} 
\newcommand{\Hop}{\hat{H}} 
\newcommand{\Hopn}{\hat{H}_{\rm 0}}
\newcommand{\Hops}{\hat{H}_{\rm s}}

\renewcommand{\vec}[1]{\mathbf{#1}}

\newcommand{\vspacebeforecaption}{\vspace{0cm}}  
\newcommand{\vspaceaftercaption}{\vspace{0cm}}  

\usepackage{xcolor}

\begin{document}

\date{\today}
\title{%
    Anderson Metal-Insulator Transitions With Classical Magnetic Impurities
}

\author{Daniel Jung}
\email{d.jung@jacobs-university.de}
\affiliation{%
    School of Engineering and Science, Jacobs University Bremen, 28759 Bremen,
    Germany
}

\author{Keith Slevin}
\email{slevin@phys.sci.osaka-u.ac.jp}
\affiliation{%
    Department of Physics, Graduate School of Science, Osaka University, 1-1
    Machikaneyama, Toyonaka, Osaka 560-0043, Japan
}

\author{Stefan Kettemann}
\email{s.kettemann@jacobs-university.de}
\affiliation{%
    School of Engineering and Science, Jacobs University Bremen, 28759 Bremen,
    Germany
}

\begin{abstract}

We study effects of classical magnetic impurities on the Anderson
metal-insulator transition numerically.  We find that a small concentration of
Heisenberg impurities enhances the critical disorder amplitude $\Wc$ with
increasing exchange coupling strength $J$.  The resulting scaling with $J$ is
analyzed which supports an anomalous scaling prediction by Wegner due to the
combined breaking of time-reversal and spin-rotational symmetry.  Moreover, we
find that the presence of magnetic impurities lowers the critical correlation
length exponent $\nu$ and enhances the multifractality parameter $\alpha_0$.
The new value of $\nu$ improves the agreement with the value measured in
experiments on the metal-insulator transition (MIT) in doped semiconductors
like phosphor-doped silicon, where a finite density of magnetic moments is
known to exist in the vicinity of the MIT.  The results are obtained by a
finite-size scaling analysis of the geometric mean of the local density of
states which is calculated by means of the kernel polynomial method.  We
establish this combination of numerical techniques as a method to obtain
critical properties of disordered systems quantitatively.

\end{abstract}

\pacs{
    71.23.An,  
    71.30.+h,  
    72.20.Ee,  
    75.20.Hr,  
    72.15.Rn,  
    64.60.an   
}

\keywords{
    Anderson localization,
    disordered solids,
    local magnetic moment,
    phase transition,
    local density of states,
    typical density of states,
    density of states,
    metal-insulator transition,
    mobility edge,
    kernel polynomial method
}

\maketitle


Experimental studies of uncompensated doped semiconductors like Si$_{1-x}$P$_x$
(Si:P) show a metal-insulator transition (MIT) as function of dopant density
$x$.  This is one of the most extensively studied cases of a quantum phase
transition \cite{Lohneysen1998,Lohneysen2000,Mott1976}.  The doping increases
the carrier density and thereby the conductivity, but also creates the onsite
disorder potential.  The random positioning of the dopants results in random
hopping amplitudes between the dopant sites.  Moreover, the electron-electron
interaction causes spin and charge correlations.  Therefore, the MIT can
neither be described completely by an Anderson MIT (AMIT)
\cite{Kramer1993,Evers2008}, driven solely by disorder, nor by a
correlation-driven MIT, the Mott transition \cite{Mott1968}.  Taking into
account both correlations and disorder remains an open problem of condensed
matter theory \cite{Lee1985,Belitz1994}.

Thermodynamic measurements prove the presence of localized magnetic moments in
the metallic regime in the excess specific heat and a low temperature
divergence of the magnetic susceptibility
\cite{Lohneysen1998,Lohneysen2000,Schlager1997,Andres1981}.  Indications of
magnetic moments can also be seen in transport measurements, such as the
thermoelectrical Seebeck coefficient \cite{Lohneysen1998,Lohneysen2000}.  These
experiments have been interpreted by assuming that up to $10\,\%$ of all
P-atoms contribute localized paramagnetic moments at the MIT
\cite{Sachdev1989,Bhatt1981,Bhatt1982,Milovanovic1989}, which originate from
the localized states in the tails of the impurity band.  Thus, it is an
essential step in understanding the MIT in doped semiconductors to understand
the influence of magnetic impurtities on the AMIT.

Since the work of Hikami et al.\ \cite{Hikami1980} it is known that weak
localization is suppressed in the presence of a finite concentration of
localized magnetic moments, because the exchange interaction with the spins of
the conduction electrons breaks time-reversal invariance.  The breaking of
time-reversal symmetry (TRS) is known to weaken Anderson localization, so the
localization length $\xi$ is enhanced.  In quasi-1D disordered wires this leads
to an enhancement $\xi = s \beta \xi_0$, with $\beta = 1$ ($\beta = 2$) when
TRS is unbroken (broken) and $s = 1$ ($s = 2$) when spin-rotational symmetry
(SRS) is unbroken (broken) \cite{Efetov1983,Kettemann2000}.  Here, $\xi_0 =
M_{\rm t} l_{\rm e}$, where $M_{\rm t}$ is the number of transverse channels
and $l_{\rm e}$ is the elastic mean free path \cite{Efetov1983}.  With SRS
intact, there are independent channels for the localization of up and down
spins. Otherwise the spin-up and spin-down channels are mixed, and the
electrons have effectively twice as many channels, which enhances $\xi$
accordingly.

Three-dimensional disordered systems are known to bear an AMIT.  It can be
expected that the breaking of TRS and SRS shifts the critical disorder $\Wc$
towards stronger disorder amplitudes $W$, which is a measure of the width of
the distribution of the disorder potential.  Likewise, the critical electron
density $n_{\rm c}$ in doped semiconductors is decreased
\cite{Khmelnitskii1981,Wegner1986}.  The symmetry class of the transition is
thereby changing from \textit{orthogonal} to \textit{unitary}
\cite{Khmelnitskii1981,Wegner1986}.  In the presence of an external magnetic
field, this change of symmetry class of the conduction electrons is governed by
the parameter $\XB = \xi^2 / \lB^2$, where $\lB$ is the magnetic length.
Therefore, in analogy, the spin scattering rate due to magnetic impurities
$\taus^{-1}$ is expected to enter through the symmetry parameter $\Xs = \xi^2 /
\Ls^2$, where $\Ls = \sqrt{\De \taus}$ is the spin relaxation length, $\De$ is
the diffusion coefficient and $\xi$ is the correlation (localization) length on
the metallic (insulating) side of the AMIT \cite{Hikami1980}.  When $\Xs \ge
1$, the electron spin relaxes before it can cover the area limited by $\xi$ and
the system is in the \textit{unitary regime}.

The crossover at the mobility edge can be studied through a scaling ansatz for
the conductivity $\sigma$ on the metallic side, as done in
Ref.~\cite{Khmelnitskii1981} for the case of an external magnetic field.
Following this approach, using the spin scattering rate $\taus^{-1}$, yields
$\sigma(\taus^{-1}) = e^2 f(\Xs)/(h \xi)$.  As function of the energy
difference $\Delta E = E - \EM$ to the mobility edge $\EM$ one then obtains
\cite{Wegner1986}
\begin{equation} \label{sigma}
    \sigma(\taus^{-1}) =
        \Delta E^{(d-2)\nu} \tilde{f}(\taus^{-1}\, \Delta E^\varphi) \quad\rm.
\end{equation}
Simple scaling theory yields $\varphi = 2\nu$ \cite{Khmelnitskii1981}, with
$\nu$ being the critical exponent describing the divergence of $\xi$ at the
mobility edge (in the 3D orthogonal universality class).  Wegner argues that
while an external magnetic field yields $\varphia = 2\nu$, the additional SRS
breaking by magnetic impurities rather yields
\begin{equation}
    \varphis = 2\nu + 3  
    \label{e:plusthree} \quad\rm,
\end{equation}
in a 2nd-order $d = 2 + \varepsilon$ expansion \cite{Wegner1986,later}.  Thus,
a numerical analysis of $\varphi$ in the presence of magnetic moments is called
for.  The value of $\varphia$ in a magnetic field has been studied in
Ref.~\cite{Drose1998}, and they find good agreement with $\varphia = 2\nu$
within their numerical accuracy for $\Wc$.  A random magnetic field should
yield the same value.  On the other hand, if a finite concentration of
classical magnetic impurities $\nM$ with spin $S$ is present, the spin
relaxation rate is finite, resulting in a shift of the critical disorder
amplitude $\Wc$ that can be obtained from the scaling ansatz for the
conductivity \eqref{sigma}.  For classical spins with a Heisenberg exchange
coupling of strength $J$, it follows
\begin{equation}
    \Wc = \Wcn + \Wcn
    \left(\frac{\ac^2}{\De \taus}\right)^{\frac{1}{\varphi}}
    \quad\text{\cite{Kettemann2012},}
    \label{e:wcj}
\end{equation}
where $\Wcn$ is the critical disorder strength without magnetic impurities,
$\De = \vF^2\, \tau / d$, $\vF$ is the Fermi velocity, and $\tau^{-1}$ is the
total elastic scattering rate. $\ac$ is a constant representing the smallest
length scale, which is identical to the lattice spacing here.

We start from the \textit{Anderson model} Hamiltonian \cite{Anderson1958b},
\begin{eqnarray}
    \Hopn =
    t\, \sum\limits_{\braket{i, j}, \sigma}\, \ket{j, \sigma} \bra{i, \sigma}
    + \sum\limits_{i, \sigma}\, V_i\, \ket{i, \sigma} \bra{i, \sigma} \quad\rm,
    \label{e:anderson}
\end{eqnarray}
where $\ket{i, \sigma}$ denotes an electron state with spin $\sigma$ located at
site $i$ of a 3D cubic lattice with $N=L^3$ sites and periodic boundary
conditions.  For the local potential $V_i$, random values are drawn from a box
distribution of width $W$, $V_i \in [-W/2, W/2]$, while the hopping amplitude
$t$ between neighboring lattice sites remains constant.

We add another term to the Hamiltonian, describing a local coupling of the
conduction electron spin $\vec{\sigma}_i$ to a classical  spin $\vec{S}_i$
(two-fluid model) \cite{Andres1981,Paalanen1988,Sachdev1989} with
$\vec{S}_i^2=S^2=1$ and random orientation (Heisenberg-like), given by the
(polar and azimuth) angles $\theta_i$ and $\varphi_i$. The angles are drawn
uniformly from the intervals $\cos\theta_i \in [-1, 1]$ and $\varphi_i \in [0,
2\pi]$.  $\vec{\sigma}_i$ are the Pauli matrices, so the general form of the
coupling term, $\sum_i J_i \vec{\sigma}_i\cdot \vec{S}_i$, can be written
\begin{multline}  
    \Hops =
    S \sum\limits_{i} J_i \Biggl(
    \cos\theta_i \sum_{\sigma=\pm 1} \sigma \ket{i, \sigma} \bra{i, \sigma} \\
    + \sin\theta_i \sum_{\sigma=\pm 1} \exp(i \sigma \varphi_i)
    \ket{i, \sigma} \bra{i, -\sigma}
    \Biggr) \quad\rm.
\end{multline}
We fix the concentration of sites carrying a magnetic moment to $\nM=5\,\%$.
Note that this is a realistic value for real materials like Si:P
\cite{Sachdev1989,Bhatt1981,Bhatt1982,Milovanovic1989}.  $J_i$ is drawn from a
binary probability distribution, $J_i \in \{J, 0\}$, taking a nonzero value
with probability $\nM$, for which it conforms to the exchange coupling strength
$J$.  Eq.~\eqref{e:anderson} by itself leads to a nonmagnetic scattering rate
$\taun^{-1} = 2\pi\, W^2\, \rho(\epsF)/(6\hbar)$.  The scattering from the
magnetic impurities enhances the total scattering rate to $\tau^{-1} =
\taun^{-1} + \taus^{-1}$.


The spin-resolved \textit{local density of states} (LDOS) is given by $\rho_{i,
\sigma}(E) = \sum_{k=1}^{2N} \left| \braket{i, \sigma|k} \right|^2
\delta(E-E_k)$, where $|k\rangle$ denotes the eigenstate with eigenenergy $E_k$
of the Hamiltonian $\Hop = \Hopn + \Hops$.  We use the \textit{kernel
polynomial method} (KPM) \cite{Weisse2006,Jung2012,Jung2014} in combination
with the \textit{Jackson kernel} which is known to yield optimal results for
the calculation of the LDOS, as it smoothens \textit{Gibb's oscillations} most
efficiently \cite{Weisse2006}.  The KPM expands the target function in a series
of \textit{Chebychev polynomials} that are only defined on the interval
$[-1,1]$, so a rescaling of the original spectrum of $\Hop$ is necessary, which
we achieve by applying a factor $a=24t$ to all energies, $E = a \Et$.  The
Jackson kernel comes with an energy broadening of $\etat = \pi / M$ at the
center of the considered interval ($E=0$), which is rescaled by the same
factor, $\eta = a \etat$ \cite{supp}.

We consider two ensemble averages: The arithmetically averaged local density of
states (ALDOS),
\begin{equation}
    \rhoav(E) =
        \frac{1}{\NS} \sum\limits_{n=1}^{\NS} \rho_n(E)\quad\rm,
    \label{aldos}
\end{equation}
which corresponds to the average density of states (ADOS) in the thermodynamic
limit (large number of samples $\NS$) \cite{ados}, and the geometrically
averaged local density of states (GLDOS),
\begin{equation}
    \rhotyp(E) =
    \exp\left(
        \frac{1}{\NS} \sum\limits_{n=1}^{\NS} \log \rho_n(E)
    \right)
    \quad\rm,
    \label{gldos}
\end{equation}
which is also known as the \textit{typical density of states}
\cite{Schubert2009}.  Here, the index $n$ takes into account both site index
$i$ and spin $\sigma$ of the conduction electrons.  $\NS$ denotes the total
number of considered local densities $\rho_n(E)$ (in this work, $\NS=8000$).
Although the LDOS is known to be spatially correlated, to save some computation
time, we take into account $p=32$ random lattice sites from each disorder
realization in the geometric mean.

\begin{figure}
    \includegraphics[width=.66\columnwidth]{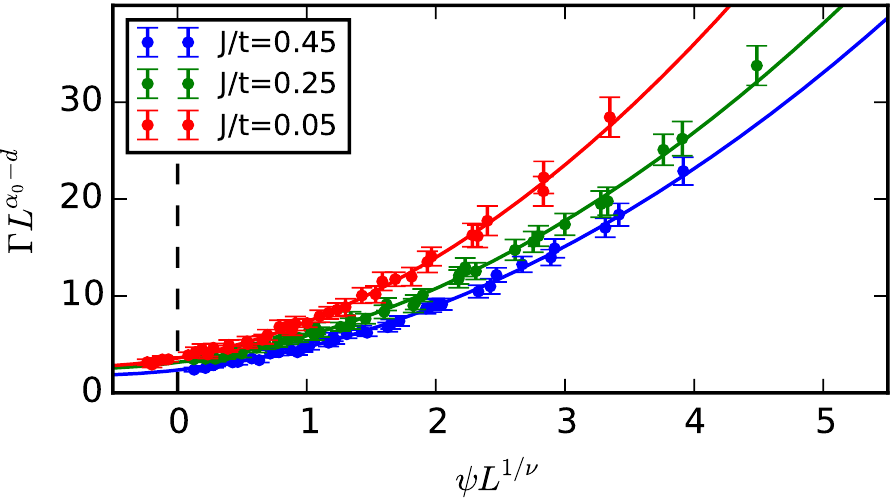}
    \vspacebeforecaption
    \caption{%
        (color online) Demonstration of the scaling ansatz \eqref{e:Gamma} at
        half filling ($E=0$) for three different values $J$ and $\nF=2$.  The
        error bars correspond to $95\,\%$ confidence.
    }
    \vspaceaftercaption
    \label{f:demoscale}
\end{figure}


In contrast to the ALDOS, the GLDOS is sensitive to the localization character
of quantum states.  It is reduced by both increasing disorder and increasing
system size within the whole energy spectrum \cite{Jung2012}.  In the
thermodynamic limit ($L \rightarrow \infty$), the GLDOS approaches zero in
energy regions of localized states, and a positive value in the case of
extended states \cite{Schubert2009}.  For finite system sizes the GLDOS stays
positive, even for perfectly localized states.  In Ref.~\cite{Schubert2009},
localised states are detected by defining a threshold value for the GLDOS,
which is adjusted to previously known values for the critical disorder $\Wc$.
For a quantitative analysis of the critical parameters (including $\Wc$), it is
necessary to perform a finite-size scaling (FSS) analysis of the GLDOS.  To
this end, we use the scaling ansatz
\begin{equation}
    \Gamma(W, L) = L^{d-\alpha_0} F(\psi L^{1/\nu}, \eta \rhoav L^d)
    \quad\text{\cite{Asada2006}}
    \label{e:Gamma}
\end{equation}
for the GLDOS at half filling ($E=0$), where $\Gamma(E) = \rhotyp(E) /
\rhoav(E)$, $d=3$, $\psi = (\Wc - W) / \Wc$ is the \emph{reduced disorder},
$\nu$ is the \emph{correlation/localization length exponent} \cite{Evers2008},
and $\alpha_0$ is a multifractality parameter
\cite{Rodriguez2011,Evers2008,Ujfalusi2015}.  We neglect the disorder
dependence of $\rhoav(E)$.  By fixing the ratio $G = L^d / M = 1$, the function
$F(x, y)$ \eqref{e:Gamma} only depends on the first argument, $x = \psi
L^{1/\nu}$.  We expand the function $\Ft(x) \equiv F(x, \rhoav \pi G)$ to order
$\nF$ in $x$, using a power series, $\Ft = \sum_{n=0}^{\nF} \Ft_n\, x^n$, so
our fit model \eqref{e:Gamma} possesses $\NP = \nF + 4$ parameters, $(\Wc,
\alpha_0, \nu, \Ft_0, \Ft_1, \cdots, \Ft_{\nF})$.  The scaling \eqref{e:Gamma}
is demonstrated in Fig.~\ref{f:demoscale} for $\nF = 2$.

Tab.~\ref{t:fit} summarizes the fit results.  Considering five system sizes $L
\in \{10, 20, \ldots, 50\}$ and eight disorder strengths $W/t \in \{15, 15.5,
\ldots, 18.5\}$, each fit has access to $\ND=40$ data points \cite{Ls}.  To
assess the quality of the fit, the $\chi^2$ statistic and the \textit{goodness
of fit} (GOF) probability $Q$ are computed \cite{supp}.  For every $J$, we
select an optimal series expansion order $\nF \in \{2, 3, 4\}$ so that $|Q -
1/2|$ is minimized, as $Q = 0.5$ indicates an optimal fit \cite{optimalvalue}.

\begin{figure}
    \begin{center}
        \subfloat{
            \label{f:fpar-wc}
            \includegraphics[width=.98\columnwidth]{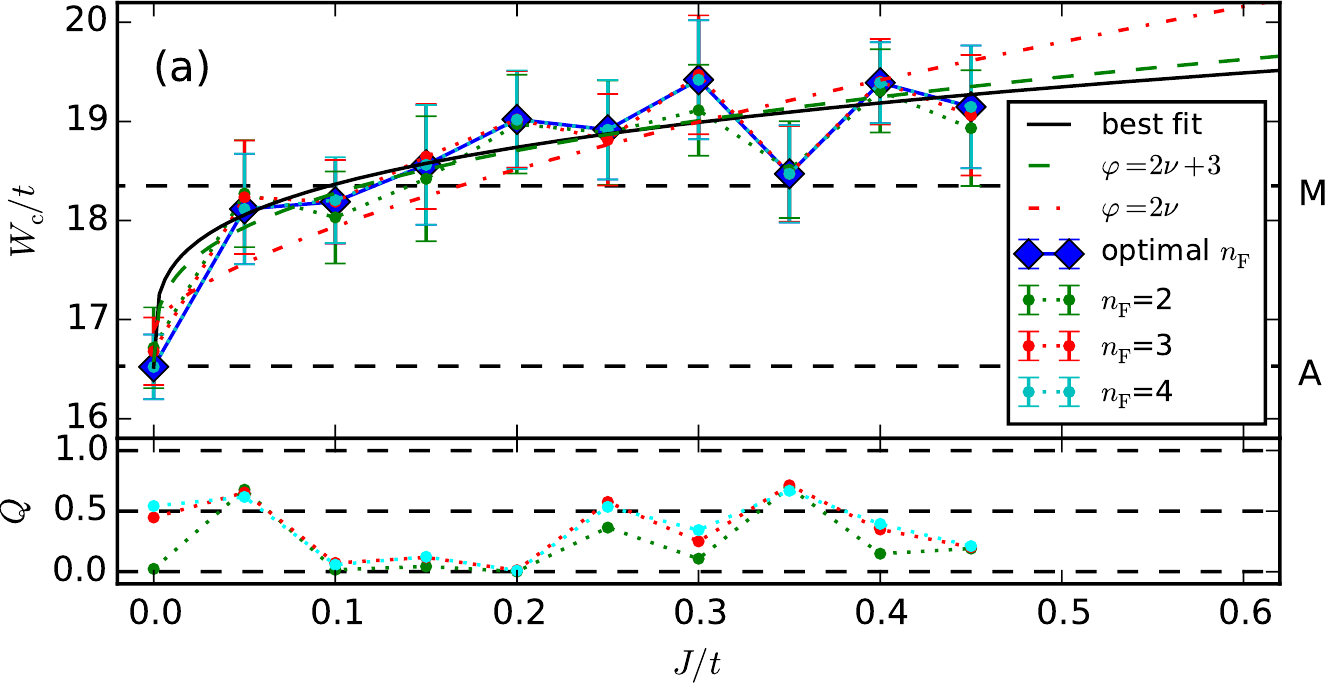}
        }\\[-2ex]
        \subfloat{
            \label{f:fpar-a0}
            \includegraphics[width=.48\columnwidth]{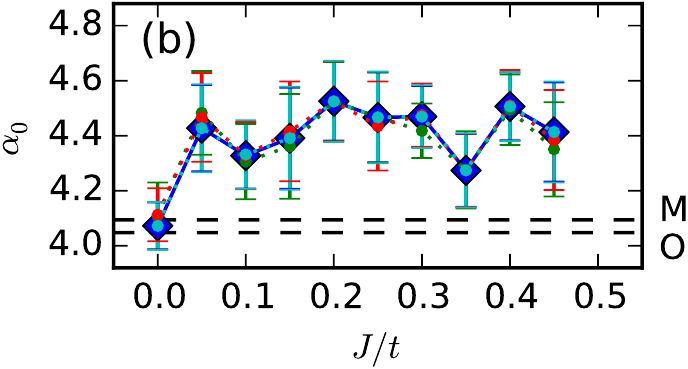}
        }
        \subfloat{
            \label{f:fpar-nu}
            \includegraphics[width=.48\columnwidth]{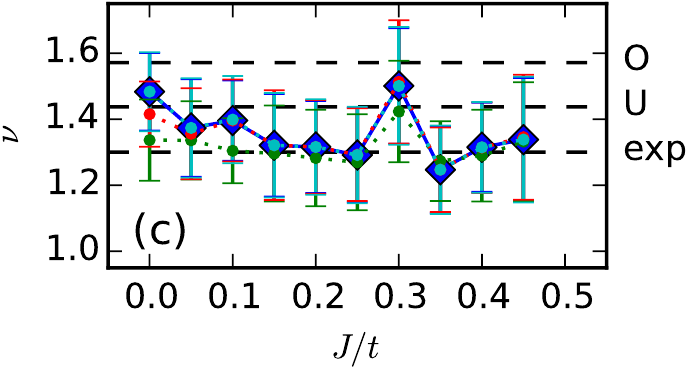}
        }
    \end{center}
    \vspacebeforecaption
    \caption{%
        (color online) Dependence of the fit parameters $\Wc$
        \protect\subref{f:fpar-wc}, $\alpha_0$ \protect\subref{f:fpar-a0} and
        $\nu$ \protect\subref{f:fpar-nu} on the exchange coupling $J$, using
        different series expansion orders $\nF$.  The dashed horizontals mark
        established values for the pure Anderson model (A, realised by our
        model for $J=0$) \cite{Rodriguez2011}, a model considering an external
        magnetic field (M) \cite{Drose1998}, the 3D orthogonal (O)
        \cite{Slevin2014,Ujfalusi2015} and the 3D unitary (U) universality
        class \cite{Ujfalusi2015}, and the experimental value (exp)
        \cite{Lohneysen1998}.  For $\Wc(J)$, the data with minimal $|Q-1/2|$
        \protect\subref{f:fpar-wc} is fitted to \eqref{e:wcjfit} using
        $\nu=1.571$ \cite{Slevin2014} (see also Tab.~\ref{t:wcj}).  The
        error bars correspond to $95\,\%$ confidence.
    }
    \vspaceaftercaption
    \label{f:fpar}
\end{figure}

Fig.~\ref{f:fpar} shows how the fit parameters $\Wc$, $\alpha_0$ and $\nu$
depend on the coupling strength $J$, using different series expansion orders
$\nF$.  Our result for $\Wc$ at $J = 0$ (corresponding to the original Anderson
model) is $\Wc^{\rm A'} / t = 16.52\pm0.17$, which agrees with established,
more precisely measured values \cite{Slevin1997,Rodriguez2011,Ujfalusi2015}
like $\Wc^{\rm A} / t = 16.530(16.524,16.536)$ \cite{Rodriguez2011}.  As the
coupling strength $J$ is increased, the critical disorder $\Wc$ approaches a
higher value of $\Wc^{\rm U}/t \approx 19.4$.  This tendency is expected, as
the symmetry class is changing from orthogonal to unitary. Another study
considering an external magnetic field has found a value of $\Wc^{\rm M} / t
\approx 18.35$ \cite{Drose1998}.  We conclude that the additional SRS breaking
of the magnetic impurities causes a further increment.  This can qualitatively
be expected, since the mixing of the spin-up and spin-down channels by the SRS
breaking enhances the number of available spin channels and thereby weakens the
localization \cite{Efetov1983,Kettemann2000}.

\begin{table}
    \caption{
        Fit parameters $\Wc$, $\alpha_0$ and $\nu$ for different exchange
        coupling strengths $J$ with their standard errors, found by expanding
        the function $\Ft(x)$ to order $\nF$.  To assess the quality of the
        fit, we provide the $\chi^2$ statistic and the goodness of fit
        probability $Q$ \cite{supp}.
    }
    \centering
    \begin{tabular}{llccccc}
        \hline\hline
        $J/t$  & $\nF$ & $\Wc/t$        & $\alpha_0$    & $\nu$         & $\chi^2$ & $Q$ \\
        \hline
        $0.00$ & $4$   & $16.52\pm0.17$ & $4.07\pm0.04$ & $1.48\pm0.06$ & $30.5$   & $0.54$ \\
        $0.05$ & $4$   & $18.12\pm0.28$ & $4.43\pm0.08$ & $1.37\pm0.08$ & $29.0$   & $0.62$ \\
        $0.10$ & $3$   & $18.19\pm0.21$ & $4.33\pm0.06$ & $1.40\pm0.06$ & $45.5$   & $0.07$ \\
        $0.15$ & $4$   & $18.56\pm0.31$ & $4.39\pm0.09$ & $1.32\pm0.08$ & $41.4$   & $0.12$ \\
        $0.20$ & $3$   & $19.02\pm0.25$ & $4.53\pm0.07$ & $1.32\pm0.07$ & $54.1$   & $0.01$ \\
        $0.25$ & $4$   & $18.92\pm0.26$ & $4.47\pm0.08$ & $1.29\pm0.07$ & $30.6$   & $0.54$ \\
        $0.30$ & $4$   & $19.42\pm0.31$ & $4.47\pm0.06$ & $1.50\pm0.09$ & $34.6$   & $0.34$ \\
        $0.35$ & $4$   & $18.47\pm0.25$ & $4.27\pm0.07$ & $1.25\pm0.07$ & $28.0$   & $0.67$ \\
        $0.40$ & $4$   & $19.39\pm0.21$ & $4.51\pm0.06$ & $1.31\pm0.07$ & $33.5$   & $0.39$ \\
        $0.45$ & $4$   & $19.15\pm0.32$ & $4.41\pm0.09$ & $1.34\pm0.10$ & $38.1$   & $0.21$ \\
        \hline\hline
    \end{tabular}
    \label{t:fit}
\end{table}

As can be seen in Fig.~\ref{f:fpar-a0}, the value of $\alpha_0$ undergoes a
gradual transition to a larger value by tuning up the coupling strength $J$.
Remarkably, this value is larger than that of a recent study using multifractal
analysis of the 3D Anderson model in a magnetic field, $\alpha_0^{\rm M} =
4.094(4.087..4.101)$ \cite{Ujfalusi2015}.  This suggests that the additional
spin symmetry breaking enhances $\alpha_0$ beyond the unitary value as obtained
when only TRS is broken. For $J = 0$, our value is in agreement with other
studies, for example $\alpha_0^{\rm O} = 4.048(4.045,4.050)$
\cite{Rodriguez2011}.

Our result for the localization length exponent $\nu$ in the orthogonal regime
($J = 0$) agrees within the achieved accuracy with established values
\cite{Slevin1997,Rodriguez2011,Slevin2014,Ujfalusi2015} like $\nu^{\rm O} =
1.571(1.563,1.579)$ \cite{Slevin2014}.  However, the error obtained within our
method is about 8 times larger than that obtained within the well-established
\emph{transfer matrix method} \cite{Slevin2014}.  During preliminary
calculations we observed that $\nu$ is increasing when lowering the ratio $G =
L^d / M$ \cite{supp}.  This is expected since $G$ is proportional to the ratio
between the KPM broadening and the average level spacing.  For $G = 0.1$ the
broadening should be of the order of the mean level spacing, while our
calculations for $G = 1$ could still mix critical and non-critical states
\cite{supp}.  Since the computational effort scales inversely linear with $G$,
we are forced to make a tradeoff between the largest considered system size $N
= L^3$, the chosen value of $G$ and the resulting computation time, so we have
decided to choose $G = 1$ for this analysis.  Note that experimental
investigations have always yielded values of $\nu$ considerably smaller than
theoretical predictions, partly because they face a similar problem of low
energy resolution \cite{Lohneysen1998}, just like our numerical method does.

For disordered systems in a magnetic field (3D unitary universality class),
values $\nu$ smaller than that of the 3D orthogonal universality class have
been reported \cite{Slevin1997,Ujfalusi2015}, like $\nu^{\rm M} =
1.437(1.426,1.448)$ \cite{Ujfalusi2015} (marked in Fig.~\ref{f:fpar-nu}).  Note
that within the achieved accuracy, our results for $\nu$ with magnetic
impurities ($J > 0$) are of similar or smaller magnitude, and in good agreement
with the experimental value $\nu^{\rm exp} \approx 1.3$ \cite{Lohneysen1998} of
real materials in which magnetic impurities are known to exist at the MIT.


The scaling of $\Wc$ with $J$ has been analyzed in Fig.~\ref{f:fpar-wc}.
Eq.~\eqref{e:wcj} suggests a scaling $\Wc(J) \sim J^{2/\varphi}$.  Hence, we
use
\begin{equation}
    \Wc(J) = a J^\mu + b
    \label{e:wcjfit}
\end{equation}
for the fit, with $\mu = 2 / \varphi$.  The fit results are summarized in
Tab.~\ref{t:wcj}.  The best fit (smallest $|Q-1/2|$) is found for Wegner's
scaling \eqref{e:plusthree} with $\epsilon = 1$ \cite{Wegner1986}.  Also the
free fit of the parameter $\mu$ shows good agreement with this analytic
prediction.  Our results do clearly not support the relation $\varphi = 2\nu$
\cite{Khmelnitskii1981,Kettemann2012}, which results in GOF probabilities that
are orders of magnitude away from an acceptable range (e.g., $Q \in
[0.1,0.9]$).  This interpretation remains intact even when using our own value
for the localization length exponent $\bar\nu = 1.48\pm0.06$ (for $J=0$)
instead of the value $\nu = 1.571(1.563,1.579)$ \cite{Slevin2014}.

\begin{table}
    \caption{
        Fit results for $\Wc(J)$. In the top row, $\mu$ is a free fit
        parameter.  Otherwise, $\mu = 2 / \varphi$ is fixed to values (shown in
        bold) according to the given analytic formulas for $\varphi$
        \cite{Wegner1987b,Kettemann2012}, using either $\nu =
        1.571(1.563,1.579)$ \cite{Slevin2014} or our own value for $J = 0$,
        $\bar\nu=1.48\pm0.06$.
    }
    \centering
    \begin{tabular}{lccccc}
    \hline\hline
    $\varphi = \ldots$ & $a$           & $\mu$           & $b$            & $\chi^2$ & $Q$ \\
    \hline
    Free fit           & $3.40\pm0.46$ & $0.27\pm0.09$   & $16.52\pm0.21$ & $11.3$ & $0.13$ \\
    \hline
    $2\nu+3$           & $3.61\pm0.34$ & $\mathbf{0.33}$ & $16.57\pm0.19$ & $12.0$ & $0.15$ \\
    $2\nu$             & $4.52\pm0.70$ & $\mathbf{0.64}$ & $16.89\pm0.26$ & $28.4$ & $4e-04$ \\
    \hline
    $2\bar\nu+3$       & $3.64\pm0.35$ & $\mathbf{0.34}$ & $16.58\pm0.19$ & $12.2$ & $0.14$ \\
    $2\bar\nu$         & $4.62\pm0.75$ & $\mathbf{0.67}$ & $16.93\pm0.27$ & $30.9$ & $1e-04$ \\
    \hline\hline
    \end{tabular}
    \label{t:wcj}
\end{table}


To conclude, we have shown numerically how local magnetic moments which break
TRS and SRS affect the metal-insulator transition in the 3D Anderson model.  We
found that the critical exponent $\nu$ decreases for increasing coupling
strength $J$ and determined its value as $\nu^{\rm S} \approx 1.3\pm0.1$ for
$5\,\%$ magnetic impurities.  Within the obtained accuracy, this value agrees
with experimental results obtained from conductivity scaling at the MIT in
phosphor-doped silicon \cite{Lohneysen1998}.  We also find the multifractality
parameter $\alpha_0^{\rm S} \approx 4.4\pm0.1$ when both TRS and SRS are
broken, a value larger than the unitary value when only TRS is broken
\cite{Ujfalusi2015}.  We considered the scaling of the critical disorder
amplitude $\Wc(J)$ and confirm an analytical prediction by Wegner
\cite{Wegner1986}.  Thus, the present investigation may relate the
systematically lower values for the critical exponent $\nu$ found in
experiments to the presence of a finite density of localised magnetic moments.
We note that magnetic moments are known to form due to the local interaction in
localised states \cite{Mott1976,Milovanovic1989}.  Recently, it has been shown
that the interplay between the Kondo screening of magnetic moments and Anderson
localization may result in a novel quantum phase transition
\cite{kondo1,kondo2,Kettemann2012}.  It remains to combine this effect with
long-range Coulomb interaction in disordered electron systems
\cite{Finkel1983,Harashima2014} in order to achieve a complete understanding of
the experimental results \cite{Lohneysen1998}.  Further, we established a
method to obtain critical properties of disordered electron systems by a
finite-size scaling ansatz for the geometric mean of the local density of
states, which enables us to make use of the kernel polynomial method to
efficiently calculate the LDOS. This method should be further developed in
order to reach an accuracy comparable to more established methods like the
transfer matrix method \cite{Slevin1999}.

\begin{acknowledgments}

We gratefully acknowledge useful discussions with Georges Bouzerar, Ki-Seok
Kim, Hyun-Yong Lee and Eduardo Mucciolo.  This research was supported by the
World Class University (WCU) program through the National Research Foundation
of Korea funded by the Ministry of Education, Science and Technology
(R31-2008-000-10059-0), Division of Advanced Materials Science.  The numerical
calculations have been performed using computational resources of the
Computational Laboratories for Analysis, Modeling and Visualization (CLAMV),
Jacobs University Bremen, Germany.

\end{acknowledgments}




\begin{thebibliography}{10}
\bibitem{Ls}
    Depending on $J$, $W$ and $L$, $\Ls < L$ may not always be fulfilled in our
    calculations \cite{supp}.
\end{thebibliography}


\end{document}